# THE MEAN FIELD APPROXIMATION FOR A SYSTEM OF TRIPLET BOSONS IN NICKELATES

V. S. Ryumshin[1], S. V. Nuzhin[1], Yu. D. Panov[1], A. S. Moskvin[1,2]

[1]Institute of Natural Sciences, Ural Federal University, Yekaterinburg, Russia

[2]M.N. Mikheev Institute of Metal Physics, Ural Branch, Russian Academy of Sciences, Yekaterinburg, Russia

E-mail: vitaliy.riumshin@urfu.ru

**Abstract**

Rare-earth orthonickelates $RNiO_3$ are Jahn-Teller magnets, unstable with respect to the anti-Jahn-Teller disproportionation reaction with the formation of a system equivalent to a system of effective spin-triplet composite bosons moving in a non-magnetic lattice. Within the framework of the two-sublattice approximation, we have developed a mean field theory for a model nickelate with competition between phases of charge ordering, an antiferromagnetic insulator and a spin-triplet superconductor, and constructed phase diagrams taking into account phase separation.

**Keywords**: rare earth orthonickelates, mean field approximation, phase separation.

**INTRODUCTION**

Orthonickelates $RNiO_3$ (R — rare earth or yttrium Y) for many years so far have been an object of intense experimental and theoretical studies [1]. The phase diagram of most rare earth nickelates comprises a metal-like disordered phase, a charge-ordered dielectric phase and an antiferromagnetic ordering phase [2]. These systems are characterized by great diversity of physical properties, including metal-insulator transitions, uncommon nature of resistivity and noncollinear magnetic structures. One exclusion here is $LaNiO_3$ which becomes an antiferromagnetic metal with decreasing temperature [3]. Some papers reported discovery of a phase coexistence (or phase separation) in nickelates Pr and Nd [2,4,5]. Unfortunately, there has been no common opinion so far regarding the formation mechanisms of nickelates' electronic structure and phase diagrams.

We started with a simple model where orthonickelates $RNiO_3$ are considered as Jahn-Teller magnetic materials which are unstable in terms of Jahn-Teller disproportionation reaction with formation of a system equivalent to the system of effective spin-triplet composite bosons moving in the non-magnetic lattice [6–9]. In the two-sublattice approximation we have developed a mean-field theory for a model nickelate which manifests the charge order, antiferromagnetic insulator, and spin-triplet superconductor phases with the phase separation effects.

## MEAN-FIELD EQUATIONS FOR A TRIPLET BOSON MODEL

Formally, the ion $Ni^{3+}$ in the low-spin configuration $t_{2g}^6 e_g^1$ of octahedron $NiO_6$ forms a Jahn-Teller center with a ground state orbital doublet $^2E$. As with many other Jahn-Teller magnets [9] the orbital degeneracy in the orthonickelates $RNiO_3$ is removed not due to a local/cooperative Jahn-Teller effect, but due to an alternative mechanism of so called anti-Jahn-Teller disproportionation [6–9]. As a result, the electronic structure of the orthonickelate becomes a formal equivalent of the system of local composite spin-triplet bosons having configuration $e_g^2$; $^3A_{2g}$, which move in the non-magnetic lattice with the $t_{2g}^6$-centers. Keeping the leading terms of the effective Hamiltonian common for such system [9], let's write down a simplified Hamiltonian for a model nickelate with a simple cubic lattice as follows:

$$\hat{H} = -t \sum_{\langle ij \rangle \gamma} \left( \hat{B}_i^{\gamma+} \hat{B}_j^{\gamma} + \hat{B}_j^{\gamma+} \hat{B}_i^{\gamma} \right) + V \sum_{\langle ij \rangle} \hat{n}_i \hat{n}_j + J \sum_{\langle ij \rangle} \hat{\sigma}_i \hat{\sigma}_j. \tag{1}$$

Here, $t$ is a transfer integral of the spin-triplet boson with preserved of spin projection $\gamma = \pm 1, 0$, $V$ - parameter of the nonlocal charge interaction, $J$ - exchange integral. Only the nearest neighbors in the simple cubic lattice are summed up. Among the other most significant simplifications, the absence of terms describing the interaction with lattice, namely with the so called "breathing" mode, shall be noted. It shall be emphasized that actual Hamiltonian (1) is a generalized version of a well-known Hamiltonian of the spinless hard-core bosons model [10] for the spin-triplet bosons case.

The basis states on the site includes four states $|n, S_\gamma\rangle$, where n — the number of bosons on the site, $S$ and $\gamma$ are value of the spin and its $z$-projection: $|1,11\rangle, |1,10\rangle, |1,1-1\rangle, |0,00\rangle$. In this basis the operators on i-site will be represented by matrices $4 \times 4$. The operator $\hat{n}_i$ is the projection to subspace $n = 1$, and the spin operators are also applied in this subspace, for instance

$$\hat{n}_i = \begin{pmatrix} 1 & 0 & 0 & 0 \\ 0 & 1 & 0 & 0 \\ 0 & 0 & 1 & 0 \\ 0 & 0 & 0 & 0 \end{pmatrix}, \quad \hat{\sigma}_{zi} = \begin{pmatrix} 1 & 0 & 0 & 0 \\ 0 & 0 & 0 & 0 \\ 0 & 0 & -1 & 0 \\ 0 & 0 & 0 & 0 \end{pmatrix}. \tag{2}$$

Operators $\hat{B}_i^{\gamma+}$ describe the boson creation in $|1,1\gamma\rangle$ state, for instance

$$\hat{B}_i^{1+} = \begin{pmatrix} 0 & 0 & 0 & 1 \\ 0 & 0 & 0 & 0 \\ 0 & 0 & 0 & 0 \\ 0 & 0 & 0 & 0 \end{pmatrix}. \tag{3}$$

It is also convenient here to introduce Cartesian components of these operators with the help of relations $\hat{B}_{xi}^\gamma = \frac{1}{2}\left(\hat{B}_i^{\gamma+} + \hat{B}_i^\gamma\right)$, $\hat{B}_{yi}^\gamma = -\frac{i}{2}\left(\hat{B}_i^{\gamma+} - \hat{B}_i^\gamma\right)$, and use vector operators $\hat{\boldsymbol{B}}_i^\gamma = \left(\hat{B}_{xi}^\gamma, \hat{B}_{yi}^\gamma\right)$.

To define the mean field approximation we use the Bogolyubov inequality to assess the system's grand potential

$$\Omega(\widehat{H}) \leq \Omega = \Psi_0 + \langle \widehat{H} - \mu \widehat{N}_b - \widehat{H}_0 \rangle, \quad \Psi_0 = -\frac{1}{\beta}\ln(\operatorname{Tr} e^{-\beta \widehat{H}_0}), \tag{4}$$

where $\mu$ — is a chemical potential, $\beta = 1/k_B T$ (further, we assume $k_B = 1$), $\widehat{N}_b = \sum_i \hat{n}_i$ — is the operator of the full bosons number, and the mean value is calculated based on the states of an ideal system with the Hamiltonian

$$\widehat{H}_0 = -\sum_{i,\gamma} f^{\gamma}_{\lambda(i)} \widehat{B}^{\gamma}_i - \sum_i \varphi_{\lambda(i)} \hat{n}_i - \sum_i g_{\lambda(i)} \hat{\sigma}_i. \tag{5}$$

We use the two-sublattice approximation, and here for the two mutually inter-penetrating lattices A and B of the simple cubic lattice $\lambda(i)$ index is introduced. The molecular fields $f^{\gamma}_{\alpha\lambda}$, $\varphi_\lambda$, $g_{\zeta\lambda}$ ($\alpha = x, y$, $\gamma = \pm 1, 0$, $\zeta = x, y, z$, $\lambda = A, B$), are the variation parameters allowing to find the best assessment for $\Omega(\widehat{H})$. Expression for $\omega = \Omega/N$, where $N$ — is the number of sites is written as follows

$$\omega = \frac{\Psi_0}{N} - zt \sum_\gamma B^{\gamma}_A B^{\gamma}_B + \frac{z}{2} V n_A n_B + \frac{z}{2} J S_A S_B$$

$$+ \frac{1}{2} \sum_\lambda \left[ \sum_\gamma f^{\gamma}_\lambda B^{\gamma}_\lambda + (\varphi_\lambda - \mu) n_\lambda + g_\lambda S_\lambda \right]. \tag{6}$$

Here, $z$ — is the number of the nearest neighbors ($z = 6$ for a simple cubic lattice), while the mean $B^{\gamma}_{\lambda(i)} = \langle \widehat{B}^{\gamma}_i \rangle$, $n_{\lambda(i)} = \langle \hat{n}_i \rangle$, and $S_{\lambda(i)} = \langle \hat{\sigma}_i \rangle$ are expressed through derivatives with respect to appropriate molecular fields, e. g. $n_\lambda = -\frac{2}{N}\frac{\partial \Psi_0}{\partial \varphi_\lambda}$. By minimizing $\omega$, we'll obtain a system of equations for molecular fields

$$\begin{cases} f^{\gamma}_\lambda = 2zt B^{\gamma}_{\bar{\lambda}}, \\ \varphi_\lambda = \mu - zV n_{\bar{\lambda}}, \\ g_\lambda = -zJ S_{\bar{\lambda}}. \end{cases} \tag{7}$$

For $\lambda$ index the overline implies a sublattice additional to this one: $\bar{A} = B, \bar{B} = A$. Equations (7) shall be supplemented with a condition for bosons concentrations $n = \langle \widehat{N}_b \rangle / N$, in order to exclude the chemical potential $\mu$: $n_A + n_B = 2n$.

Equations (7) for high temperature have solutions which describe the non-ordered (NO) phase, where $B^{\gamma}_\lambda = 0$, $S_\lambda = 0$ и $n_A = n_B = n$. In this case an analytical expression for the free energy per lattice site can be formulated

$$f = \frac{z}{2} V n^2 - \frac{1}{\beta}[n \ln 3 - n \ln n - (1-n)\ln(1-n)]. \tag{8}$$

The expression in brackets defines the entropy of NO phase with maximum equal to $2\ln 2$ at $n = 3/4$.

The conditions of the grand potential minimum's instability for NO phase relative to variations of fields $\varphi_\lambda$, $\boldsymbol{g}_\lambda$ or $\boldsymbol{f}_\lambda^\gamma$ result in expressions for critical temperatures of transition into the charge-ordered (CO) phase with $x = (n_A - n_B)/2 \neq 0$, antiferromagnetic (AFM) phase with $\boldsymbol{g}_A = -\boldsymbol{g}_B \neq 0$ or boson superfluid (BS) phase $\boldsymbol{B}_\lambda^\gamma \neq 0$, respectively

$$T_{CO} = zVn(1-n), \qquad (9)$$

$$T_{AFM} = \frac{4}{3}zt\left(n - \frac{3}{4}\right)\left[\ln\frac{n}{3(1-n)}\right]^{-1}, \qquad (10)$$

$$T_{BS} = \frac{2}{3}zJn. \qquad (11)$$

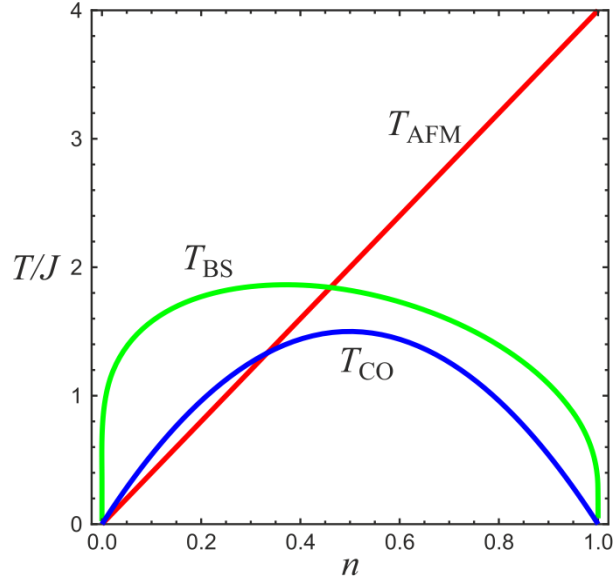

Figure 1. Concentration dependencies of critical temperatures (9−11) at z = 6, V = 1, J = 1, t = 1.

Concentration dependencies of critical temperatures are shown in Figure 1. Expression (9) for $T_{CO}$ coincides with the well-known expression for critical temperature of the charge-ordering in the model of hard-core bosons [10]. The difference of expression (10) for $T_{BS}$ from the superfluid transit critical temperature [10] is associated with the spin degeneracy for bosons in our model. The linear concentration dependence (11) for critical temperature of the antiferromagnetic ordering is itself a specific ratio which is evidenced by comparison with results for other pseudospin models [11,12]. If the temperatures are below the critical one the system of equations (7) has solutions which can be called pure phases when only one order parameter is nonzero. For CO phase there remain molecular fields equations $\varphi_\lambda$, $\lambda = A, B$:

$$\varphi_\lambda = \mu - zV\frac{3}{3 + e^{-\beta\varphi_{\bar{\lambda}}}}. \qquad (12)$$

For AFM phase, given the isotropic nature of exchange interaction, it is enough to consider the molecular fields equations $\varphi = \varphi_A = \varphi_B$ and $g = g_{zA} = -g_{zB}$:

$$\begin{cases} \varphi = \mu - zV\dfrac{1 + 2\text{ch}(\beta g)}{1 + 2\text{ch}(\beta g) + e^{-\beta\varphi}}, \\ g = zJ\dfrac{2\text{sh}(\beta g)}{1 + 2\text{ch}(\beta g) + e^{-\beta\varphi}}. \end{cases} \quad (13)$$

For BS phase it is helpful to consider that boson transport in the Hamiltonian (1) has isotropic nature and doesn't depend on the spin projection. In this case it is enough to determine the molecular fields $\varphi = \varphi_A = \varphi_B$ and $f = f_{xA}^\gamma = f_{xB}^\gamma$, $\gamma = A, B$, taking all other fields equal to zero. In this case the system of equations (7) will be written as

$$\begin{cases} \varphi = \mu - zV\dfrac{\phi\left[2e^{\frac{\beta}{2}\varphi} + \text{ch}\left(\frac{\beta}{2}\phi\right)\right] + \varphi\text{sh}\left(\frac{\beta}{2}\phi\right)}{2\phi\left[e^{\frac{\beta}{2}\varphi} + \text{ch}\left(\frac{\beta}{2}\phi\right)\right]}, \\ 1 = zJ\dfrac{\text{sh}\left(\frac{\beta}{2}\phi\right)}{\phi\left[e^{\frac{\beta}{2}\varphi} + \text{ch}\left(\frac{\beta}{2}\phi\right)\right]}, \end{cases} \quad (14)$$

where $\phi = \sqrt{\varphi^2 + 3f^2}$. It should be emphasized that obtained solutions for pure phases should be verified for stability.

Generally speaking, apart from the pure phases, solutions for mixed phases can be implemented under certain conditions, when several order parameters are non-zero. These solutions are akin to the supersolid phase in the hard-core bosons model [10]. Investigation of properties and availability of these types of solutions is quite a timeconsuming task. However, our tentative results demonstrate that free energy of mixed phases in the considered model with the given set of parameters is higher than free energy of the phase separation state of pure phases. This situation is fully equivalent to results for the model of hard-core bosons [13]. The phase separation areas in the phase diagram are defined using Maxwell construction [14]: at given temperature the boundary concentrations $n_i$ corresponding to pure phases $i = 1,2$ can be found from the concentration ratios of chemical potentials of phases: $\mu_i(n, T) = \mu^*$, where $\mu^*$ — point of intersection of specific grand potentials of phases, $\omega_1(\mu^*, T) = \omega_2(\mu^*, T)$. The ratios $m_i$ of each phase are given by the following relations:

$$m_1 = \frac{n_2 - n}{n_2 - n_1}, \quad m_2 = \frac{n - n_1}{n_2 - n_1}, \quad n_1 \leq n \leq n_2. \quad (15)$$

**RESULTS**

Typical phase diagrams for the model of spin-triplet bosons with Hamiltonian (1) obtained with regard to the approximations above are given in Figure 2.

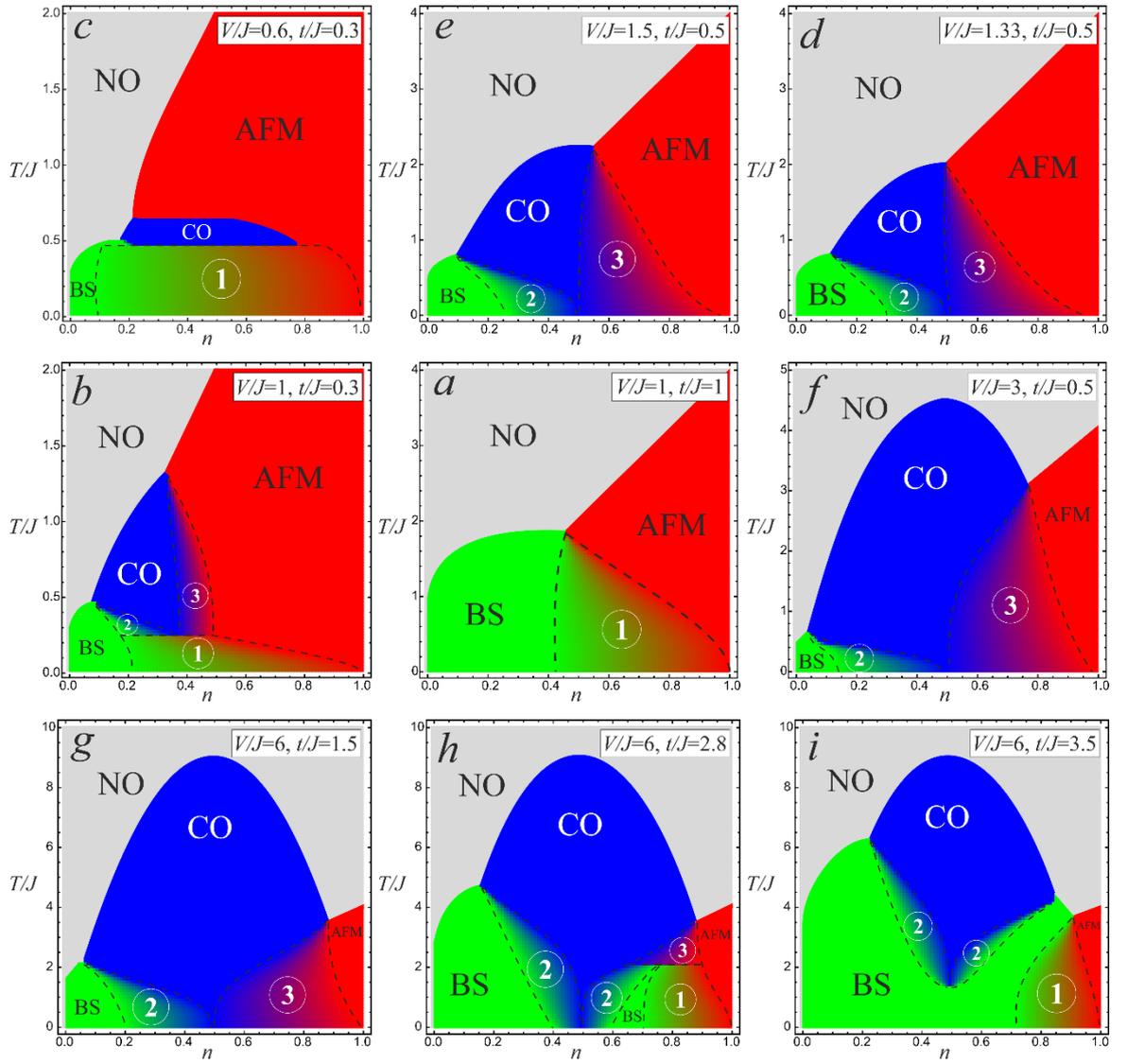

Figure 2. Phase diagrams of triplet bosons model with Hamiltonian (1) in the mean field approximation. Dashed lines show the boundaries of phase separation indicated by numbers: 1 — BS+AFM, 2 — BS+CO, 3 — CO+AFM.

At $V/J = 1$, $t/J = 1$ (see Figure 2, a) the phase diagram has the areas of BS and AFM phases divided by the phase separation area BS+AFM denoted by 1 in a circle. This area starts in the tricritical point where critical temperatures of BS and AFM phases intersect, and in its ground state this area covers the range $0.42 < n < 1$. When the transfer integral is reduced to a value $t/J = 0.3$ (see Figure 2, b) we arrive at proportional decrease of $T_{BS}$, while CO phase area existing only at finite temperatures appears on the phase diagram. Two new tricritical points represent the initial areas of the phase separation BS+CO and CO+AFM (denoted by 2 and 3 in a circle), which are also possible only at finite temperatures. In the ground state the phase separation BS+AFM in $0.22 < n < 1$ range takes place. Decrease of the charge interaction parameter to $V/J = 0.6$ (see Figure 2, c) leads to reduction of the CO phase area and removal of the phase separation areas

BS+CO and CO+AFM, while BS+AFM area, in contrary, increases and covers the range $0.1 < n < 1$ in its ground state.

At $V/J = 1.33$, $t/J = 0.5$ (see Figure 2, d) the tricritical point at the intersection of $T_{CO}$ and $T_{AFM}$ corresponds to $n = 0.5$. In the ground state in point $n = 0.5$ the CO phase is provided, yet the boundary of the phase separation area CO+AFM at finite temperatures lies slightly more to the left of this point. In ground state BS+CO and CO+AFM areas cover the regions $0.3 < n < 0.5$ and $0.5 < n < 1$, respectively. Increase of $V$ parameter, as shown in Figure 2, e and f, results in proportional increase of $T_{CO}$ and offset of tricritical points, however, in its ground state the area CO+AFM remains the same, $0.5 < n < 1$, and the ranges BS and BS+CO are reallocated in the range $0 < n < 0.5$.

The step-by-step change of the phase diagram at $V/J = 6$ with the growth of transfer integral from $t/J = 1.5$ to $t/J = 3.5$ is shown in Figure 2, g, h, i. When $t/J = 2.8$ the area of BS phase and BS+CO and BS+AFM areas appear from the right to the point $n = 0.5$, however, at this point in the ground state, the CO phase is realized. Further, when $t/J = 3.5$ the CO phase exists only at finite temperatures, and in ground state only BS phase and phase separation BS+AFM take place.

For the ground state of the considered model within the applied approximation several common conclusions may be formulated. Given small values $n$ the ground state of the system is BS phase, CO phase appears only for relatively large values of parameter V only when $n = 0.5$, and AFM phase takes place only when $n = 1$. Finite ranges of values n may correspond to phase states BS, BS+AFM, BS+CO and CO+AFM.

It should be noted also that the applied approximation didn't include the solutions for mixed phases like super solid and its equivalents when more than one parameter of the order is non-zero. Finding these solutions may be an individual research issue, however, our preliminary results demonstrate that similar to the case of the well-known model of spinless hard-core bosons [13] the free energy of mixed phases in our model is higher than the free energy of the phase separation state of pure phases.

**CONCLUSION**

Within the mean field approximation we have considered a simple model of the disproportionate phase of orthonickelates RNiO3, the Hamiltonian of which is equivalent to the Hamiltonian of the spin-triplet local bosons model on a nonmagnetic lattice. Phase diagrams of coexistence of "pure" phases with a single non-zero order parameter have been constructed — namely, the phase of non-magnetic charge ordering, the antiferromagnetic phase and the spin-triplet bosonic superconductor phase. A universal behavior of phase separation is demonstrated.

**FUNDING**

This study was supported by the Ministry of Science and Higher Education of the Russian Federation, project FEUZ-2023-0017.